\documentclass{INTERSPEECH2023}

\usepackage{CJK}
\usepackage{cite}
\usepackage{amsmath}
\usepackage{cases}
\usepackage{multirow}
\usepackage{verbatim}

\title{A Systematic Exploration of Joint-training for Singing Voice Synthesis}

\name{
    Yuning Wu$^{1}$, 
    Yifeng Yu$^{2}$,
    Jiatong Shi$^{3}$, 
    Tao Qian$^{1}$, 
    Qin Jin$^{1*}$\thanks{*Corresponding author.}
}
\address{
    $^{1}$ Renmin University of China,
    $^{2}$ Georgia Institute of Technology,
	$^{3}$ Carnegie Mellon University}
 \email{\{yuningwu, qiantao, qjin\}@ruc.edu.cn, yyu479@gatech.edu, jiatongs@andrew.cmu.edu}

\begin{document}

\maketitle
 
\begin{abstract}
There has been a growing interest in using end-to-end acoustic models for singing voice synthesis (SVS). Typically, these models require an additional vocoder to transform the generated acoustic features into the final waveform. However, since the acoustic model and the vocoder are not jointly optimized, a gap can exist between the two models, leading to suboptimal performance. 
Although a similar problem has been addressed in the TTS systems by joint-training or by replacing acoustic features with a latent representation, adopting corresponding approaches to SVS is not an easy task. How to improve the joint-training of SVS systems has not been well explored. In this paper, we conduct a systematic investigation of how to better perform a joint-training of an acoustic model and a vocoder for SVS. We carry out extensive experiments and demonstrate that our joint-training strategy outperforms baselines, achieving more stable performance across different datasets while also increasing the interpretability of the entire framework.  
\end{abstract}

\section{Introduction}
\label{sec:intro}
\begin{figure}[t]
	\centering
	\includegraphics[width=1.0\columnwidth]{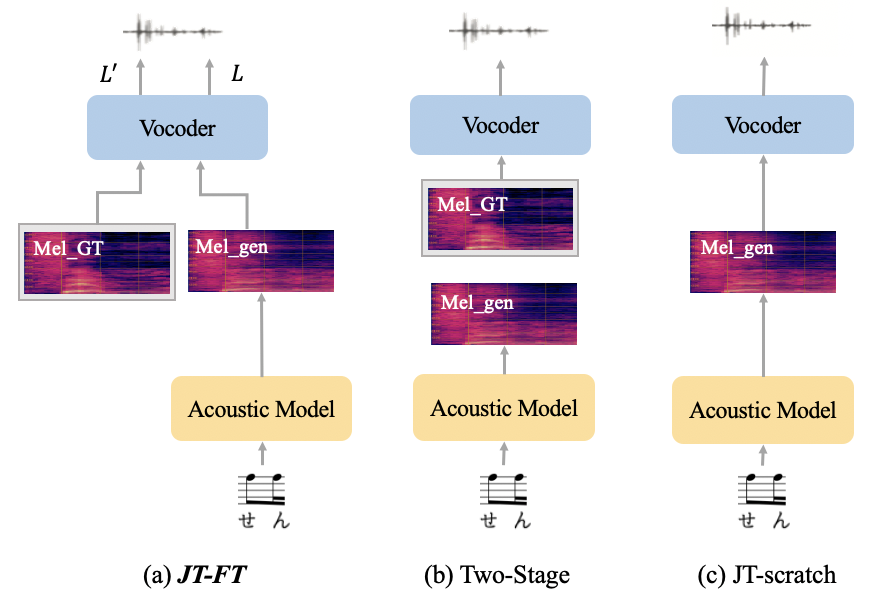}
	\caption{\small Implementations of training in different SVS systems. }
	\label{fig:model}
\end{figure}

Singing voice synthesis (SVS) is an engaging area of research that aims to generate a singing voice from a given musical score, which includes information such as lyrics, pitch, and note duration~\cite{cook1996singing, hono2021sinsy, kenmochi2007vocaloid}. In recent years, the development of deep learning has led to significant advances in SVS systems. Typically, state-of-the-art systems for SVS consist of two core components: an acoustic model \cite{hono2021sinsy, zhang2022susing, nishimura2016singing, chen2020hifisinger, lu2020xiaoicesing, ren2020deepsinger, hono2019singing, liu2022diffsinger, guo2022singaug, baseline2, kim2018korean, zhang2022wesinger, zhuang2021litesing, tae2021mlp, lee2021n, gu2021bytesing} that converts the music score into acoustic features and a vocoder \cite{kong2020hifi, huang2022singgan, huang2021multi, refinegan, morise2016world} that transforms the acoustic features into waveforms. These two components are usually trained in a two-stage fashion (Fig.~\ref{fig:model}(b)), with each being optimized independently. One of the most commonly used acoustic features in these systems is the Mel spectrograms.

However, there is a problem with this conventional two-stage SVS system: the disparity in acoustic features between the training and inference stages \cite{VISinger}. During training, the vocoder is trained on ground truth acoustic features extracted from raw singing data. However, during inference, the vocoder is provided with generated acoustic features, which can lead to degraded performance. 
A straightforward approach to address this problem is to jointly train the acoustic model and vocoder~\cite{VISinger} (Fig.~\ref{fig:model}(c)). 
However, such an approach can be challenging as the vocoder is trained on the imperfect acoustic feature predictions from the acoustic model, thus providing noisy gradient information to the acoustic model. This makes it difficult for both the acoustic model and the vocoder to converge well. We therefore often observe performance degradation with such direct joint-training.
Moreover, it becomes even more challenging for the joint-training when considering the limited availability of public singing datasets due to copyright restrictions and the high cost of annotation \cite{baseline2, guo2022singaug, ren2020deepsinger}. 
Furthermore, in addition to achieving high accuracy and naturalness in pronunciation, SVS systems are expected to capture the variability of melody and master rhythm \cite{liu2022diffsinger, chen2020hifisinger, liu2021vibrato}. These difficulties for acoustic modeling can further increase the uncertainty for the vocoder. Despite these challenges, there has been a lack of investigation into how to improve the joint-training of SVS systems. 

On a similar synthesis task, text-to-speech (TTS), the problem of the acoustic feature gap with Mel spectrograms has been addressed by  joint-training \cite{Lim2022JETSJT, Hayashi2021ESPnet2TTSET, weiss2021wave} or by replacing the acoustic feature with a latent representation, such as in the DelightfulTTS2 \cite{Delightfultts2} and  VITS \cite{VITS} systems. 
However, adopting these approaches to SVS is not an easy task, as it must comply with both the pitch and duration constraints of the musical score. 
In this paper, we carry out a systematic investigation of how to better perform a joint-training of an acoustic model and a vocoder for SVS. Specifically, we use spectrogram-based acoustic features as intermediate representations in our joint-training SVS system to overcome potential issues with latent representations. We systematically compare different joint-training strategies on different SVS benchmarks and summarize some key factors that are essential for the quality of the synthesized singing based on joint-training. We finally provide an optimal joint-training strategy for SVS (Fig.~\ref{fig:model}(a)), and demonstrate that it outperforms other solutions on different benchmarks. 
The contributions of this work include:


\begin{itemize}
\item We conduct a systematic investigation of various joint-training approaches for SVS with spectral acoustic features.
\item We summarize some key factors that are essential for the quality of the synthesized singing based on joint-training. 
\item We provide an optimal joint-training strategy that outperforms both the conventional two-stage and the latent-representation-based systems.
\end{itemize}

\section{Methodology}
\label{sec:method}



Our joint-training SVS framework consists of an acoustic model and a vocoder, which can freely adopt any model. We only show an example selection for the acoustic model and vocoder in the following experiments.

The acoustic model is composed of an encoder-decoder structure connected by a duration predictor that predicts the acoustic frame lengths of the phonemes, which are then expanded by a length regulator to match the extracted acoustic features from the raw singing. In this paper, we choose the commonly-used Mel spectrogram as the intermediate acoustic feature for its advantageous interpretability. 
The vocoder is responsible for converting the acoustic feature into singing waveforms. We use the state-of-the-art HiFi-GAN vocoder \cite{kong2020hifi} because of its high-fidelity synthesis quality. HiFi-GAN is a vocoder with Generative Adversarial Network (GAN) architecture that consists of a generator and two discriminators: a multi-period discriminator and a multi-scale discriminator, which capture local and global information, respectively.

\begin{figure}[t]
	\centering
	\includegraphics[width=0.85\columnwidth]{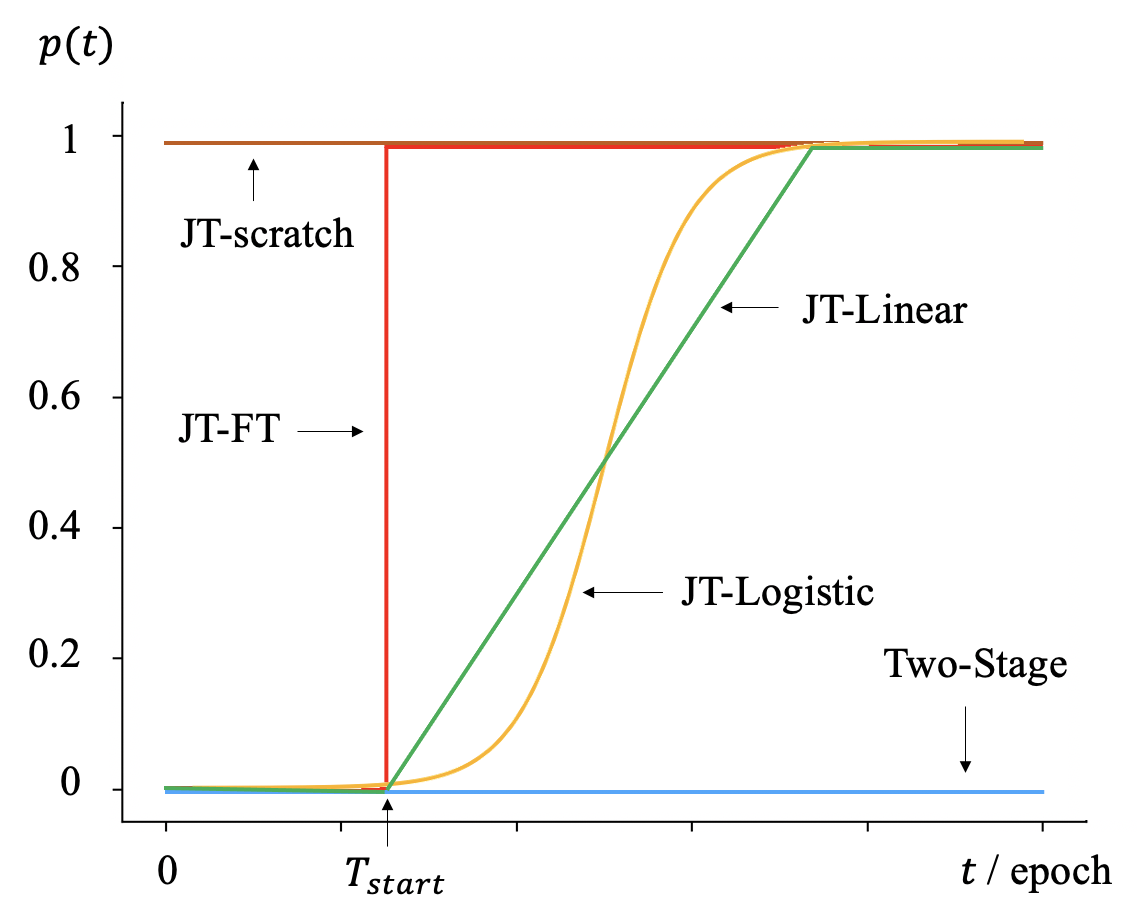}
	\caption{\small Curves of $p(t)$. Different curves correspond to different growth patterns as discussed in Sec.~\ref{sec:method}.}
	\label{fig:curve}
\end{figure}

\subsection{Systematic Exploration of Joint-Training}
\label{ssec: mixture of intermediate acoustic feature}


As discussed in Section~\ref{sec:intro}, it is challenging for the vocoder to converge on the imperfect Mel spectrograms. 
We therefore adopt a scheduled sampling strategy to train the vocoder in order to mitigate the gap between the acoustic model and the vocoder. Besides, using ground truth features can help the vocoder converge more quickly.
Specifically, we train the vocoder using a combination of acoustic features generated by the acoustic model and ground truth features extracted from the target singing waveforms. During training, we weigh the contribution of the generated acoustic features and the ground truth with weights $p(t)$ and $1-p(t)$ respectively. 
During the training, as the acoustic model is not well trained at the early stage, the ground truth acoustic features should be assigned higher weight, while as the training progresses, the predicted acoustic features from the acoustic model become more reliable and can be assigned increasing weights $p(t)$.
Therefore, properly setting $p(t)$ will impact joint-training difficulty.




According to general scheduled sampling strategies, we can make two assumptions of $p(t)$:
\begin{itemize}
    \item $p(t)$ is a monotonically increasing function
    \item the increasing pattern can be described by a growth rate $r$
\end{itemize}
We therefore formulate $p(t)$ according to the logistic function\footnote{\url{https://en.wikipedia.org/wiki/Logistic_function}} as follows:
\begin{subnumcases}
{p(t)=}
    0,&  $t \in [0, \leq T_{\text{start}}$),\\
    \frac{Ke^{rt}}{K+e^{rt}-1}, & $ t \in [T_{\text{start}}, T_{\text{end}}$),\\
    1,&  $t \in [T_{\text{end}},T_{\text{max}})$,
\end{subnumcases}
where $t$ denotes the current training epoch, and $T_{\text{max}}$ represents the maximum training epoch. 
[$T_{\text{start}}$, $T_{\text{end}}$] refers to the training period where the ground truth acoustic features and predicted acoustic features are both utilized as supervision with weights of $1-p(t)$ and $p(t)$. 
$K \in [0, 1]$, and $r \in \mathbb{R}$ are two hyper-parameters that affect the joint-training performance:
\begin{itemize}
    \item $K$: Final proportion ratio. This parameter represents the final weight of the $p(t)$. If $K$ is 0, the training pipeline degenerates into a two-stage system without dealing with the gap between the acoustic model and the vocoder. If $K$ is 1, it becomes a joint-training system that directly passes the predicted acoustic feature to the vocoder.
    \item $r$: Changing rate. This parameter measures the growth patterns and the shape of the $p(t)$ curve. If $r$ is 0, the pipeline is equivalent to a finetuned system where the SVS system is jointly finetuned with a pretrained acoustic model and a pretrained vocoder. 
\end{itemize}

\subsection{Losses }
\label{ssec: losses}
We define $x$ as the ground truth Mel-spectrogram, $\hat{x}$ as the predicted Mel-spectrogram from the acoustic model, $w$ as the ground truth waveform, and $\hat{w}$ as the predicted waveform from the vocoder. 

As the ground truth Mel spectrogram extracted from saw singing is leveraged in the joint-training, 
the total loss of the joint-training SVS system consists of an acoustic model loss $L_{\text{AM}}$ and a vocoder loss $L_{\text{v}}$ and details are as follows:


The acoustic model loss $L_{\text{AM}}$ is calculated with a Mel spectrogram loss $L_{m_a}$ and a duration loss $L_{d}$ for the duration predictor.
        \begin{equation}
        L_{\text{AM}} = \lambda_{d} L_{d} + \lambda_{m_a} L_{m_a},
        \end{equation}
where we set $\lambda_{d}$, and $\lambda_{m_a}$ to 1.

The vocoder loss $L_{\text{v}}$ consists of 
a adversarial loss $L_{\text{adv}_\text{mix}}$, a feature matching loss $L_{f_\text{mix}}$, a Mel spectrogram loss and a discriminator loss $L_{D_{\text{mix}}}$ :
\begin{itemize}
    \item The adversarial loss $L_{\text{adv}_\text{mix}}$ of the generator in the vocoder.
        \begin{equation}
        L_{\text{adv}}(x)=\mathbb{E}_{w}\Big[(\textit{D}(\textit{G}(w)-1)^2\Big],
        \end{equation}
        
        \begin{equation}
        L_{\text{adv}_{\text{mix}}}=p(t) L_{\text{adv}}(\hat{w}) + (1-p(t)) L_{\text{adv}}(w).
        \end{equation}
    \item The feature matching loss $L_{f}$, which measures the distance between the generated singing waveforms and the ground truth.
        \begin{equation}
        L_{f_{\text{mix}}}= p(t) L_{f}(\hat{w}) + (1-p(t)) L_{f}(w).
        \end{equation}
    \item The Mel spectrogram loss $L_{m_s}$, which is extracted from the synthesized singing and the ground truth singing signals.

        \begin{equation}
        L_{m_{\text{mix}}}= p(t) L_{m_s}(\hat{x}) + (1-p(t)) L_{m_s}(\varphi(w)),
        \end{equation}
        where $\varphi$ is the procedure used to convert singing waveforms to Mel spectrograms.

    \item The $D$ denote the discriminator in HiFi-GAN and the loss of discriminator $L_D$ is defined as the same as 
        \begin{equation}
            L_D(x)=\mathbb{E}_{w,x}\Big[(\textit{D}(w)-1)^2+(\textit{D}(\textit{G}(x))^2\Big],
        \end{equation}
        \begin{equation}
            L_{D_{\text{mix}}} = p(t) L_D(\hat{x}) + (1-p(t)) L_D(x).
        \end{equation}
\end{itemize}




The formulation of vocoder loss  $L_{\text{v}}$ is as follows: 
\begin{equation}
L_{\text{v}}= \lambda_{\text{adv}} L_{\text{adv}_{\text{mix}}} + \lambda_{f} L_{f_{\text{mix}}} + \lambda_{m} L_{m_{\text{mix}}},
\end{equation}

\begin{equation}
L_G=L_{\text{AM}} + \lambda_{\text{adv}} L_{\text{adv}_{\text{mix}}} + \lambda_{f} L_{f_{\text{mix}}} + \lambda_{m} L_{m_{\text{mix}}},
\end{equation}
where we set $\lambda_{\text{adv}}$ to 1, $\lambda_{f} = 2$, and $\lambda_{m} = 45$.

The final training loss $L_{\text{tot}}$ of the joint-training SVS system is:

\begin{equation}
L_{\text{tot}}= L_{\text{AM}} + L_{\text{v}},
\end{equation}


\section{Experiment}
\label{sec:exp}

\subsection{Experimental Settings}
\label{ssec: exp settings}

\begin{table}[t]
\caption{ \small Comparison of the best implementation of joint training system, \textbf{\textit{JT-FT}}, with the two-stage system and VISinger on Ofuton. The joint training system is tested on a Transformer based acoustic model with HiFi-GAN vocoder. Evaluations include four objective metrics (\textbf{MCD}, \textbf{F\_0 RMSE}, \textbf{VUV\_E}, and \textbf{SA}) and a subjective metric (\textbf{MOS}) are described in Section~\ref{ssec: exp settings}. * indicates significant improvement of MOS over all variants with p-value$<$0.005.}

\resizebox{1\linewidth}{!}{
\centering
 \small
\begin{tabular}{c|lccccc}
\toprule
\textbf{Dataset}          & \textbf{Model}     & \textbf{MCD ↓} & \textbf{F\_0 RMSE ↓}   & \textbf{VUV E ↓}      & \textbf{SA ↑} & \textbf{MOS ↑} \\ 
\midrule
\multirow{4.5}{*}{Ofuton}   & Two-Stage    & 6.60   & \textbf{0.09} & 2.25\%    & \textbf{63.62}\%   & 3.54±0.06    \\
                          & \textit{\textbf{JT-FT}}   & 6.39  & 0.10   & 2.92\%     & 63.61\%   & \textbf{3.87}±0.06*   \\
\cmidrule(l){2-7}
                          & VISinger       & \textbf{6.30}  & \textbf{0.09}    & \textbf{2.24}\%          & 59.08\%         &  3.48±0.07    \\
\cmidrule(l){2-7}
                          & Ground Truth     & -    & -     & -           & -        &  4.66±0.04    \\

\bottomrule
\end{tabular}

\label{tab: main result}
}
\end{table}

\begin{table}[!t]
\caption{ \small Supplementary experiments of the comparisons between \textbf{\textit{JT-FT}} and two-stage system. Some of results of Ofuton are listed in Table~\ref{tab: main result}.}
\centering
 \resizebox{1\linewidth}{!}{
\begin{tabular}{c|lcccc}
\toprule
\textbf{Dataset}          & \textbf{Model}     & \textbf{MCD ↓} & \textbf{F\_0 RMSE ↓}   & \textbf{VUV E ↓}      & \textbf{SA ↑}  \\ 
\midrule
\multirow{2}{*}{Ofuton}   & LSTM-Two-Stage    & 6.22  & 0.10  & 2.54\%   & 60.82\%    \\
                           & LSTM-\textit{\textbf{JT-FT}}  & 6.19 & 0.10    & 2.79\%     & 61.54\%     \\
\midrule
\multirow{4}{*}{Opencpop} & LSTM-Two-Stage    & 7.76  & 0.21 & \textbf{8.40\%}      & 57.57\%     \\
                          & LSTM-\textit{\textbf{JT-FT}}     & \textbf{7.59} & 0.23    & 8.68\%     & \textbf{60.51\%}       \\
                          & Transformer-Two-Stage   & 7.72  & 0.19 & 9.31\%     & 60.27\%      \\
                          & Transformer-\textit{\textbf{JT-FT}}            & 7.72    & \textbf{0.18}     &   9.19\%          & 60.27\%       \\
\bottomrule
\end{tabular}
}
\label{tab: xxx}
\end{table}

\noindent\textbf{\textit{Dataset.}}
we conduct SVS experiments on two public datasets: Ofuton-P \cite{Ofuton} and Opencpop \cite{Opencpop}. The Ofuton-P dataset is composed of 56 Japanese pop songs and has a singing voice recording duration of around 1 hour. The Opencpop dataset contains 100 Chinese pop songs and has a singing voice recording duration of about 5.2 hours.

\noindent\textbf{\textit{Model Architecture.}}
Our joint-training SVS system is based on the ESPnet2-Muskits \footnote{\url{https://github.com/espnet/espnet}} \cite{shi2022muskits, watanabe2018espnet} with two core components: the acoustic models and the vocoder. Specifically, we employ a Transformer-based acoustic model following \cite{lu2020xiaoicesing}. The supplementary experients also test a three-layer 256-dimensional bidirectional Long-Short Term Memory units (Bi-LSTM) based acoustic model from \cite{RNN}. We use HiFi GAN as the vocoder \footnote{\url{https://github.com/kan-bayashi/ParallelWaveGAN}} \cite{kong2020hifi}. We also implement VISinger \cite{VISinger} for comparison.

\noindent \textbf{\textit{Training.}}
We train our joint-training SVS system with both Transformer-HiFi GAN and LSTM-HiFi GAN models using the HiFi GAN architecture with a sampling rate of 24 kHz and a batch size of 16. Our training is performed with a maximum of 500 epochs and 500 iterations per epoch, with the training seed set to 777. We employ an AdamW optimizer with a learning rate of 1.25e-5 and a weight decay of 0.0. For both the generator and discriminator, we use an exponential learning rate scheduler with a gamma value of 0.999875.

\noindent\textbf{\textit{Evaluation Metric.}}
We evaluate our joint training SVS model using several objective and subjective metrics. For objective evaluation, we use Mel Cepstral Distortion (MCD) and F0 Root Mean Square Error (RMSE) to measure the similarity between the generated singing voice and the ground truth. For subjective evaluation, we conduct a Mean Opinion Score (MOS) test where participants rate the quality of the generated singing voices on a scale from 1 (bad) to 5 (excellent). Each audio is graded by 40 different participants. We conduct subjective evaluations on all joint-training systems with the Transformer-based acoustic model and HiFi GAN vocoder on the Ofuton dataset and report the MOS with 95\% confidence interval.

\begin{table}[t]
\caption{\small Ablation of the performance of final proportion ratio.}
\centering
\resizebox{1\columnwidth}{!}{
\begin{tabular}{lccccc}
\toprule
\multicolumn{1}{c}{\textbf{Value}}       & \textbf{MCD ↓}    & \textbf{F\_0 RMSE ↓}     & \textbf{VUV\_E ↓}       & \textbf{SA ↑}    & \textbf{MOS ↑}    \\
\midrule
K=0  (JT-\textit{scratch})              & 7.89        & 0.22            & 2.53\%      & 47.73\%            & 2.32±0.06 \\
\midrule
K=0.25                & 9.07  & 0.26    & 3.53\%   & 41.51\% & 1.91±0.05 \\
K=0.5              & 9.29          & 0.28            & 5.69\%      & 34.63\%            & 1.81±0.05 \\
K=0.75                & 9.23  & 0.27   & 2.97\%               & 34.39\%   & 1.94±0.06 \\
\midrule
K=1  (Two-Stage)    &  \textbf{6.60} & \textbf{0.09} & \textbf{2.25\%}   & \textbf{63.62\%}    & \textbf{3.54}±0.06* \\
\bottomrule
\end{tabular}
}
\label{tab: K}
\end{table}

\begin{table}[t]
\caption{\small Ablation of the performance of changing rate.}
\centering
\resizebox{1\columnwidth}{!}{
\begin{tabular}{lccccc}
\toprule
\multicolumn{1}{c}{\textbf{Pattern of r}}       & \textbf{MCD ↓}    & \textbf{F\_0 RMSE ↓}     & \textbf{VUV\_E ↓}       & \textbf{SA ↑}    & \textbf{MOS ↑}    \\
\midrule
JT-Linear                & 8.56  & 0.23   & 3.50\%                & 36.28\%   & 1.93±0.05 \\
JT-Logistic              & 8.31    & 0.27       & 5.17\%         & 37.18\%            & 1.85±0.05 \\
\textbf{\textit{JT-FT}}               & \textbf{6.39}  & \textbf{0.10}   & 2.92\%               & \textbf{63.61\%}   & \textbf{3.87}±0.06* \\
\midrule
JT-scratch    & 7.89        & 0.22            & \textbf{2.53\%}      & 47.73\%            & 2.32±0.06 \\
\bottomrule
\end{tabular}
}
\label{tab: r}
\end{table}

\begin{table}[t]
\caption{\small Ablation of the performance of $T_{\text{start}}$. (ME is the number of max\_epoch)}
\centering
\resizebox{1\columnwidth}{!}{
\begin{tabular}{lccccc}
\toprule
\multicolumn{1}{c}{\textbf{Value}}       & \textbf{MCD ↓}    & \textbf{F\_0 RMSE ↓}     & \textbf{VUV\_E ↓}       & \textbf{SA ↑}    & \textbf{MOS ↑}    \\
\midrule
0 ME (Two-Stage)         & 6.60   & \textbf{0.09} & 2.25\%    & 63.62\%   & 3.54±0.06   \\
\midrule
0.25 ME               & \textbf{6.39}  & 0.10   & 2.92\%               & 63.61\%   & \textbf{3.87}±0.06 \\
0.5  ME             & 6.44           & 0.10            & \textbf{1.89\%}      & \textbf{63.76\%}            & 3.73±0.06 \\
0.75 ME               & 6.43  & 0.10   & 1.94\%               & 63.76\%   & 3.83±0.06 \\
\midrule
1 ME (JT-\textit{scratch})       & 7.89        & 0.22            & 2.53\%      & 47.73\%            & 2.32±0.06 \\
\bottomrule
\end{tabular}
}
\label{tab: P}
\end{table}

\begin{table}[t]
\caption{\small Analysis study of Mel-spectrograms gap.}
\centering
\resizebox{1\columnwidth}{!}{
\begin{tabular}{lccccc}
\toprule
\multicolumn{1}{c}{\textbf{Method}}       & \textbf{MCD ↓}    & \textbf{F\_0 RMSE ↓}     & \textbf{VUV\_E ↓}       & \textbf{SA ↑}    & \textbf{MOS ↑}    \\
\midrule
Two-Stage     & 6.60   & \textbf{0.09} & 2.25\%    & 63.62\%   & 3.54±0.06   \\
Cascade                & \textbf{6.53}  & 0.10   & \textbf{2.13\%}               & \textbf{63.77\%}   & \textbf{3.72}±0.06* \\
\midrule
JT-scratch      & 7.89        & 0.22            & 2.53\%      & 47.73\%            & 2.32±0.06 \\
\bottomrule
\end{tabular}
}
\label{tab: analysis}
\end{table}

\subsection{Comparison with baselines}
\label{ssec: comparison with baselines}

We compare the best joint-training SVS system with a two-stage system and latent-representation-based end-to-end system VISinger. Here, the best joint training SVS system is equivalent to jointly finetuning a pretrained acoustic model and a pretrained vocoder. The process of exploring the best system will be illustrated in Section~\ref{ssec: ablation study}. We conduct the experiment on both Ofuton and Opencpop datasets with Bi-LSTM-based or Transformer-based encoder-decoder structure acoustic models.

As shown in Section~\ref{tab: main result}, the best joint-training system outperforms the conventional two-stage system on both subjective and objective metrics. The best model can not only eliminate the Mel-spectrogram gap between training and inference but also get a higher MOS than VISinger. 

\subsection{Ablation Study}
\label{ssec: ablation study}

We carry out ablation studies to investigate some factors that could influence the joint-training performance. 

\noindent
\textbf{Ablation of the Final Proportion Ratio $K$}
\label{sssec: comparison of final proportion ratio}

We compare the performance with different values of the final mixing ratio. Table~\ref{tab: K} shows that as the proportion increases, the performance first drops and then improves. This trend suggests that if the vocoder receives both the generated and ground truth Mel-spectrograms until the end of training, it cannot converge on the output of the acoustic model. The vocoder requires sufficient time to reach convergence with only the generated feature as input. Therefore, it is preferable to set the final proportion ratio to 1 or 0. As $K$=$0$ corresponds to the two-stage system with Mel-spectrograms gap between training and inference, we set the final value of proportion $K$ as 1.

\noindent
\textbf{Ablation of the Changing Rate $r$}
\label{sssec: ablation of the changing rate}

We set the final proportion $K$ to 1 and investigate different ways to increase the percentage of generated Mel-spectrograms. The growth patterns of $p(t)$ are described by the changing rate $r$, and the visualization curves of different patterns can be found in Fig~\ref{fig:curve}. The results in Table~\ref{tab: r} indicate that the vertical growth system outperforms the linear and logistic growth systems. However, these systems still perform worse than the model trained entirely on generated features. This finding further supports that the vocoder cannot effectively handle both the generated Mel-spectrograms and the ground truth as inputs, and thus cannot converge on both distributions simultaneously. Notably, the vertical growth system refers to a specific case of joint training implementation, where a pretrained acoustic model and a pretrained vocoder are jointly fine-tuned. The following ablation analysis will focus on this joint fine-tuning system.

\noindent
\textbf{Ablation of the $T_{\text{start}}$ }
\label{sssec: ablation of the translation period}

Building on the insights gained from the above ablation studies, we investigate the $T_{\text{start}}$ in the joint fine-tuning system, where the transition interval indicates the time to switch from ground truth to generated Mel-spectrograms. Interestingly, Table~\ref{tab: P} shows that the fine-tuning system can achieve relatively good performance when the switch occurs at either one quarter or three quarters of the total training time. We conjecture that without a limitation on training time, the fine-tuned systems could achieve even better performance.

\subsection{More Analysis}
\label{ssec: analysis study}

To investigate the importance of Mel-spectrograms in the SVS system as discussed in the introduction, we train two vocoders separately: one using ground truth Mel-spectrograms and the other using the output of a fully-pretrained acoustic model. Surprisingly, the vocoder trained on generated Mel-spectrograms outperforms the one trained on ground truth Mel-spectrograms. The difference in performance between the two generated waveforms highlights the existence of the Mel-spectrogram gap in training and inference of the two-stage SVS system. However, directly connecting the acoustic model and vocoder through joint-training proves challenging, as shown by the poor performance in Table~\ref{tab: analysis}. This is because that it is difficult for the vocoder to learn from the changeable intermediate feature. Nevertheless, training the vocoder on the generated acoustic feature alone yields inferior results compared to the joint fine-tuning system, indicating that joint training can also improve the acoustic model.

\begin{figure}[t]
	\centering
	\includegraphics[width=1.0\columnwidth]{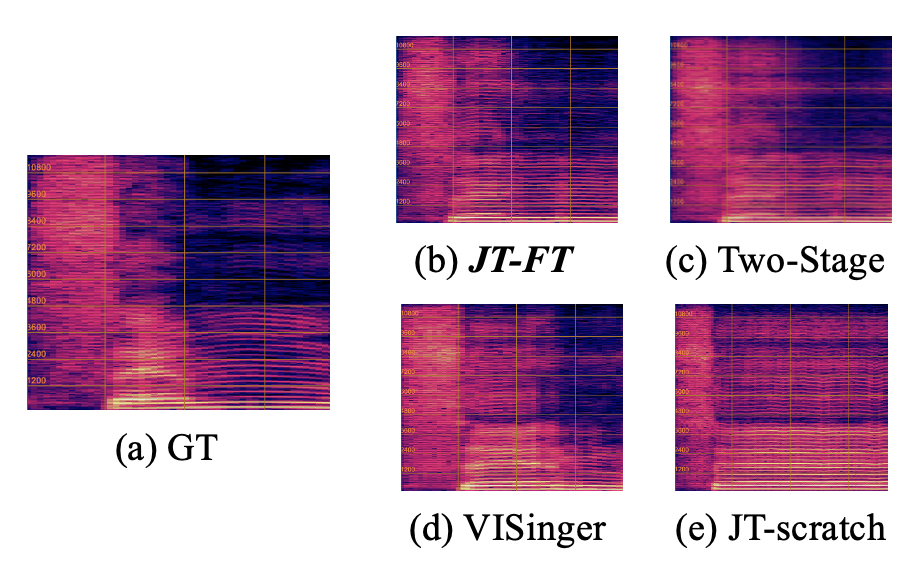}
	\caption{\small The cases of Mel spectrograms generated by various SVS systems.  Detailed discussion is in Sec.~\ref{ssec: analysis study}.}
	\label{fig:case}
\end{figure}

The Mel spectrograms generated by various SVS systems are presented in Fig \ref{fig:case}. The fine-tuning system outperforms the baselines, particularly in high-frequency bands. It is noteworthy that the joint-training from scratch adds extra noise to the signals, which aligns with our findings in Sec.~\ref{ssec: ablation study}.

\section{Conclusion}

To address the acoustic feature gap in conventional two-stage SVS systems, we investigate different implementations of joint-training SVS systems by connecting the acoustic model and the vocoder. 
We summarize some key factors that impact the joint-training performance. We provide an optimal joint-training strategy for SVS, and demonstrate that it outperforms the conventional two-stage system and the latent-representation-based VISinger. The limitation of our study is that we did not conduct experiments on large SVS datasets to test the impact of larger datasets on the joint-training SVS system. Nevertheless, we believe that extending our investigation to further explore joint-training for other tasks would be promising.

\bibliographystyle{IEEEtran}
\bibliography{refs}

\end{document}